\documentclass[12pt]{article}
\begin{document}
\title{{\bf Singular spacetime\\ and \\quantum probe}}

\author{ W{\l}odzimierz Piechocki\thanks{E-mail address: piech@fuw.edu.pl}\\
So{\l}tan Institute for Nuclear Studies, \\Ho\.{z}a 69, 
00-681 Warszawa, Poland }

\date{}
\maketitle

\date{}
\maketitle

\begin{abstract}

We examine the dependence of quantization on global properties of 
a classical system. Quantization based on local properties  may lead to 
ambiguities and inconsistency between local and global symmetries of 
a quantum system. Our quantization method based on global characteristics 
has sound foundation. Presented results give insight into the nature 
of removable type singularity of spacetime at the quantum level. 


\end{abstract}

\section{Introduction} 

Finding a theory of quantum gravity is not only an intellectual 
adventure but also the present day need, since the number of cosmological 
data on the very early Universe  increases and they call for theoretical 
description. It seems that understanding of the nature of spacetime 
singularities in quantum context is the core of the problem. The insight 
can be achieved by studying some  suitable `toy models' which include 
both singular spacetimes and quantum rules.

The toy model considered here concerns a spacetime with removable 
type singularities. We analyze classical and quantum dynamics of a test 
particle in singular and corresponding regular spacetimes. To quantize 
classical dynamics of the systems we apply the group quantization method [1]. 
We show that quantization of the system with regular spacetime can be carried 
out without problems. In the singular case not all classical observables are 
well defined globally since spacetime includes incomplete geodesics. 
Therefore, smaller number of classical observables can be mapped into quantum 
observables.  As the result, the classical dynamics of a test particle in 
a spacetime with removable type singularities can be quantized, but the local 
symmetry of the quantum singular system cannot be as rich as in the 
corresponding regular case. 

\section{Classical dynamics}

We present dynamics of a test particle in two de Sitter's type spacetimes.
For simplicity we restrict ourselves to two dimensional 
spacetimes. In conclusions we make comments concerning the four 
dimensional cases. 

The considered spacetimes are defined to be
\begin{equation}
V_p = (R^1\times R^1,~\hat{g}),~~~~~~~~~V_h = (R^1\times S^1,~\hat{g}).
\end{equation}
In both cases  the metric tensor $g_{\mu\nu} :=(\hat{g})_{\mu\nu}~~ 
(\mu,\nu =0,1)$ is defined by the line-element
\begin{equation}
ds^2 = dt^2 - exp(2t/r)~dx^2,
\end{equation}
where $r$ is a positive real constant.

\noindent
Eq.(1) presents all possible topologies of the de Sitter spacetime in two 
dimensions.
$V_p$ is a plane with global $(t,x)\in R^2$ 
coordinates. $V_h$ is defined to be a one-sheet 
hyperboloid embedded in 3d Minkowski space. There exists [3] an isometric 
immersion map $f$ of $V_p$ into $V_h$ defined by
\begin{equation}
 V_{p} \ni (t,x)\longrightarrow f(t,x):=(y^0,y^1,y^2) \in V_h,
 \end{equation}
where
$$ y^0=r\sinh (t/r) +\frac{x^2}{2r}~\exp(t/r),$$
$$ y^1=-r\cosh (t/r)+\frac{x^2}{2r}~\exp(t/r),~~~~
 y^2= -x \exp(t/r),$$
and where 
\begin{equation}
 (y^2)^2+(y^1)^2-(y^0)^2=r^2.
\end{equation}
 One can check that $f$ maps $V_p$ onto a non-compact half of $V_h$ 
 and that the induced metric on $V_h$ is identical to the metric 
defined by (2).

Let us emphasize that $f$ cannot be extended to an isometry $F:V_p 
\rightarrow V_h$ such that $F(V_p)=V_h-\{x(R^1)\}=:V_x,$ where $x:R^1 
\rightarrow V_h$ is defined to be a complete smooth curve on $V_h,$ 
and such that $V_x$ is simply connected [4]. In other words, $V_p$ and 
$V_h$ are not almost globally isometric spacetimes having only different 
boundary conditions.

It is known [5] that $V_p$ is geodesically incomplete. But all 
incomplete geodesics in $V_p$ can be extended to complete ones in $V_h$, 
which means that $V_p$ has removable type singularities. The singularity type 
of $V_p$ is not as severe as in the case of spacetimes with essential type 
singularities [6]. The latter includes both incomplete geodesics and blowing up 
curvature scalars. In our case both  $V_p$ and $V_h$ have constant 
curvatures $R= -2r^{-2}$.
 
The action integral, $S$, describing a  relativistic test particle of 
mass $m$  in gravitational field  $g_{\mu \nu}~ (\mu,\nu =0,1)$ is 
proportional to the length of a particle world-line 
\begin{equation}
S=\int \;L(\tau)~d\tau,~~~~~L(\tau):=-m\sqrt{g_{\mu\nu}(x^0(\tau),x^1(\tau))\;
\dot{x}^\mu (\tau)\dot{x}^\nu (\tau)},
\end{equation}
where $\tau$ is an evolution parameter, $x^\mu$ are 
spacetime coordinates and $\dot{x}^\mu := dx^\mu/d\tau $. It is assumed 
that $ \dot{x}^0 >0 $, i.e. $x^0$ has interpretation of time monotonically 
increasing with $\tau$. 

The Lagrangian  (5) is invariant under the reparametrization
$ \tau\rightarrow f(\tau)$.  
This gauge symmetry leads to the constraint 
\begin{equation}
 G:= g^{\mu\nu}p_\mu p_\nu -m^2=0,
\end{equation} 
where $g^{\mu\nu}$ is an inverse of $g_{\mu\nu}$ and  $p_\mu := 
\partial L/\partial\dot{x}^\mu$ are canonical momenta. 

Since a test particle does not modify the geometry of 
spacetime, the local symmetry of the system is defined by the set of all 
Killing vectors of  spacetime. Each Killing vector field 
$X^\mu~ (\mu =0,1)$ has corresponding dynamical integral [7] 
\begin{equation} 
D=p_\mu X^\mu . 
\end{equation}

\noindent 
The dynamical integrals  restricted to the constraint
surface (6) and specifying  particle trajectories admissible by the dynamics 
define the physical phase-space $\Gamma$. 

Since the hyperboloid (4) is invariant under the proper 
Lorentz transformations,  the symmetry group of $ V_h $ is $SO^{\uparrow}(1,2)$.
In the standard parametrization [2]  the infinitesimal 
transformations of $SO^{\uparrow}(1,2)$ group read   

\begin{eqnarray}
(\rho,~\theta)\longrightarrow(\rho,~\theta+a_0 r),~~~~~~~~~~~~~~~~~~~~
\nonumber \\
(\rho,~\theta)\longrightarrow(\rho-a_1 r \sin \rho /r~\sin 
\theta /r,~\theta+a_1 r\cos \rho /r~\cos \theta /r),\nonumber \\
(\rho,~\theta)\longrightarrow(\rho+a_2 r\sin \rho /r~\cos 
\theta /r,~\theta+a_2 r\cos \rho /r~\sin \theta /r),
\end{eqnarray}
where $(a_0,a_1, a_2) \in R^3 $ are parameters. 

\noindent
The corresponding dynamical integrals  (7)
\begin{eqnarray}
J_0=p_{\theta}~r,~~~~~J_1=-p_\rho ~r\sin \rho /r~\sin 
\theta /r + p_\theta ~r\cos \rho /r~\cos \theta /r, \nonumber \\
J_2= p_\rho ~r\sin \rho /r~\cos \theta /r 
+ p_\theta ~r\cos \rho /r~\sin \theta /r,~~~~~~~~~~~~
\end{eqnarray}
(where $p_\theta :=\partial L/\partial\dot{\theta},~p_\rho :=
\partial L/\partial\dot{\rho}$ are canonical momenta)
satisfy the commutation relations of $sl(2,R)$ algebra
\begin{equation}
\{ J_a , J_b \} =\varepsilon_{abc}\eta^{cd} J_d ,
\end{equation}
where $\varepsilon_{abc}$
is the anti-symmetric tensor with $\varepsilon_{012}=1$ and $ \eta^{cd} $ 
is the Minkowski metric tensor.

The constraint (6) in terms of (9) reads
\begin{equation}
J_0^2 - J_1^2 - J_2^2 = -\kappa^2,~~~~~~\kappa =mr .
\end{equation}

\noindent
The particle trajectories are found [2] to be defined by 
\begin{equation}
J_a y^a =0,~~~~~J_2 y^1 - J_1 y^2 =r^2 p_\rho .
\end{equation}
Each point point $(J_0, J_1, J_2)$ of the one-sheet hyperboloid (11) defines 
a particle trajectory (12) available for dynamics.
Therefore (11) defines the physical phase-space $\Gamma_h$ with $SO^{\uparrow}(1,2)$ 
as the symmetry group. The spacetime and phase-space of $V_h$ system 
have the same symmetry group.  

The infinitesimal symmetry transformations of $V_p$ system read [2]
\begin{equation}
(t,~x)\longrightarrow (t,~x+b_0),~~~~~
(t,~x)\longrightarrow (t-rb_1,~x+xb_1 ) ,
\end{equation}
\begin{equation}
(t,~x)\longrightarrow (t-2rxb_2 ,~x+ (x^2 +r^2 e^{-2t/r})b_2),
\end{equation}
where $(b_0,b_1,b_2) \in R^3 $ are parameters.

\noindent
The corresponding dynamical integrals (7) are
\begin{equation}
 P=p_x,~~~K=-rp_t+xp_x,~~~M=-2rxp_t +(x^2 + r^2 e^{-2t/r}) p_x ,
\end{equation}
where $p_x = \partial L/\partial\dot{x},~ p_t = \partial L/\partial \dot{t}$. 

\noindent
One can check that the integrals (15) satisfy the commutation relations of 
$sl(2,R)$ algebra in the form
\begin{equation}
\{P,K\}=P,
\end{equation}
\begin{equation}
\{K,M\}=M,~~~\{P,M\}=2K.
\end{equation}
The constraint (6) leads to 
\begin{equation}
K^2-PM=\kappa^2 .
\end{equation}
In  case of $V_p$ system, contrary to the case of $V_h$ system, some points 
$(P,K,M)$ of (18) cannot describe the trajectories available for a particle: 

\noindent
For $P=0$ there are two lines $ K=\pm \kappa$ on the hyperboloid (18).
Since by assumption $\dot{t} >0$, we have that $p_t = \partial L/
\partial\dot{t}=-m\dot{t}\:(\dot{t}-\dot{x}\exp(2t/r))^{-1/2} < 0 $. 
According to (15) $K-xP = -rp_t,$ thus $K-xP > 0,~$ i.e. $K>0~$ for $P=0$.
Therefore, the line $(P=0,~K=-\kappa)$ is not available for the dynamics. 
The hyperboloid (18) without this line defines the physical phase-space 
$\Gamma_p$. 
The particle trajectories are [2] 
\begin{equation}
x=M/2K,~~~~~~~~~~~~~~~~~~~~~~~~~~\mbox{for}~~~~P=0
\end{equation}
and 
\begin{equation}
xP=K-\sqrt{\kappa^2 +(rP)^2\exp{(-2t/r)}},~~~~~\mbox{for}~~~~P\neq 0 .  
\end{equation}
Since $\Gamma_p$ is topologically equivalent to a plane $R^2$ we can 
parametrize $\Gamma_p$ as follows [2] 
\begin{equation}
P=p,~~~K=pq-\kappa,~~~M=pq^2 -2\kappa q ,
\end{equation}
where $(q,p)\in R^2$.

The local symmetry of both $V_p$ and $V_h$ systems is defined by  
$sl(2,R)$ algebra. However, the symmetry groups are different. The Lie group 
$SO^{\uparrow}(1,2)$ cannot be the global symmetry of $V_p$ system, since $V_p$ is only 
a subspace of $V_h$ due to (3). Since the Killing vector field generated by 
(14) is not complete (see, App.A of [2] ), the  symmetry group of $V_p$ is 
the Lie group with the Lie algebra defined by (16) only. Therefore, in case 
of $V_p$ system the Lie algebra corresponding to the symmetry group is 
different from $sl(2,R)$ algebra of all the Killing vector fields. 

The  classical observables are defined to have the following properties: 

\noindent
(i) they specify particle trajectories available for dynamics ($V_h$ and $V_p$ 
are integrable systems), (ii) they are gauge invariant (have vanishing 
Poisson's brackets with the constraint $G$, Eq.(6)), (iii) they satisfy the 
algebra corresponding to the symmetry group of $V_h$ or $V_p$ system {\bf or} 
$(\widetilde{iii})$ they satisfy the algebra corresponding to  the local 
symmetry of $V_h$ or $V_p$ system.

\section{Quantum dynamics} 

In case of $V_h$ system the classical observables 
are $J_0, J_1 $ and $J_2$. We choose the following  parametrization [2] 
\begin{equation}
J_0=J,~~~J_1=J\cos\beta - \kappa\sin\beta,~~~J_2=-J\sin\beta 
-\kappa\cos\beta ,
\end{equation}
where $J\in R^1$ and $\beta\in S^1 $ are the canonical coordinates.

\noindent
The observables satisfy both sets of conditions 
$A:=\{(i),(ii),(iii)\}$ and $\tilde{A}:=\{(i),(ii),(\widetilde{iii}) \}$, 
since $sl(2,R)$ is the algebra of $SO^{\uparrow}(1,2)$ group.

The quantum observables corresponding to (22) read  [2] 
\begin{equation}
\hat{J}_0=\frac{\hbar}{i}\frac{d}{d\beta},~~~
{\hat{J}}_1 = \cos \beta 
~\hat{J_0} - ({\kappa} -\frac{i\hbar}{2})\sin \beta,~~~
\hat{J}_2 =  -\sin \beta 
~\hat{J_0} - ({\kappa} -\frac{i\hbar}{2})\cos \beta .
\end{equation}
Since they are unbounded, they are defined [2] on a dense subspace $\Omega$ 
of the Hilbert space $L^2[0,2\pi]$ 
\begin{equation}
\Omega := \{\psi\in L^2[0,2\pi]~|~\psi\in C^\infty[0,2\pi],~\psi^{(n)}(0)
= \psi^{(n)}(2\pi),~ n=0,1,2...\} .
\end{equation}
Eqs. (23) and (24) define an essentially self-adjoint representation of 
$sl(2,R)$ algebra (see, App.C of [2] ). It can be further examined for its integrability 
to the unitary representation of $SO^{\uparrow}(1,2)$ group.

In case of $V_p$ system the two sets of conditions $A$ and $\tilde{A}$ lead 
to different sets of observables, because $SO^{\uparrow}(1,2)$ is no longer the symmetry 
group of the system. We examine the consequences of each 
choice separately. 

With the choice $\tilde{A}$ the classical observables are $P,K$ and $M$. 
The symplectic transformation $(q,p)\rightarrow (I,\sigma)$ defined by 
\begin{equation}
q:=-\cot\frac{\sigma}{2},~~~p:=(1-\cos\sigma)(I+\kappa\cot\frac{\sigma}{2})
\end{equation}
leads to
\[ I_0:=\frac{1}{2}(M+P)=I,\]
\begin{equation}
I_1:=\frac{1}{2}(M-P)=I\cos\sigma -\kappa\sin\sigma,
~~~~~I_2:=K=-I\sin\sigma -\kappa\cos\sigma.
\end{equation}
where $I\in R^1 $ and $0<\sigma<2\pi $ are the canonical coordinates.

\noindent
One can easily check that the line $(P=0, K=-\kappa)$ of the hyperboloid (18) 
turns into the generatrix $(I_0=I_1, I_2 =-\kappa)$ of the hyperboloid (11) 
(with $J_a$ replaced by $I_a$).

\noindent 
The observables $I_a~(a=0,1,2)$ satisfy $sl(2,R)$ algebra, have the same 
functional form as $J_a~(a=0,1,2)$ observables, but are defined on different 
domains. Therefore, the functional form of corresponding quantum observables 
$\hat{I_a}$ is again defined by (23) (with $\hat{J_a}$ replaced by 
$\hat{I_a}$ and $\beta$ replaced by $\sigma$). However, the carrier space 
of $\hat{I_a}$ is different. Now, it is defined to be [2]
\begin{equation}
\Omega_\alpha:=\{\psi\in L^2[0,2\pi]~|~\psi\in C^\infty [0,2\pi],~\psi^{(n)}
(0)=e^{i\alpha}\psi^{(n)}(2\pi),~n=0,1,2,...\},
\end{equation}
where $0\leq \alpha <2\pi$. 

\noindent
It can be easily proved (see, App.C of [2] ) that Eq. (23), with $\hat{J_a}$ 
replaced by $\hat{I_a}$ and $\beta$ replaced by $\sigma$, and Eq. (27) 
define  essentially self-adjoint representations of $sl(2,R)$ algebra. 
However, the choice $\tilde{A}$  leads to ambiguities. 
Since the end points of the range of $\sigma$ in (26) do not coincide, 
there is no reason to choose any specific value for $\alpha$. Thus, there 
are infinitely many unitarily nonequivalent quantum systems corresponding 
to a single classical $V_p$ system. Such a quantum theory has no 
predictability. It can be useful as a phenomenological model, but we are 
interested in finding a fundamental description.
Obviously, these representations cannot be lifted to the unitary 
representation of the symmetry group, because the symmetry group has the 
algebra defined by Eq.(16) which is only a subalgebra of $sl(2,R)$ algebra. 

Now, we consider the choice $A$. The set of observables consists of $P$ 
and $K$ which satisfy the algebra (16). 
The corresponding quantum observables read [2]
\begin{equation}
\hat{P}=\frac{\hbar}{i}\frac{d}{dq},~~~~\hat{K}=q\hat{P} 
+ \frac{\hbar}{2i}-\kappa .
\end{equation}
The common invariant dense domain  $\Lambda$ for $\hat{P}$ and $\hat{K}$ 
is defined to be [2]
\begin{equation}
\Lambda := \{~\psi\in L^2 (R)~|~\psi\in C_0^\infty (R)~\}.
\end{equation}
The representation defined by (28) and (29) is  essentially self-adjoint 
(see, App.D of [2] ). Next step would be lifting this representation 
 to the unitary representation of the symmetry group of $V_p$ system. 

\section{Conclusions} 

It is interesting that at the phase-space level one has 
$\Gamma_h =\Gamma _p \cup \{$generatrix$\}$, which is quite different from 
the correspondence between $V_h$ and $V_p$ systems at the spacetime level. 
This subtle phase-space difference between $V_p$ and $V_h$ systems leads 
to quite different  quantum systems. The problem of incomplete 
geodesics of $V_p$ spacetime translates into the boundary condition problem 
at the quantum level. The latter leads to ambiguities. Uniqueness of the 
quantization procedure can be achieved by favouring the global symmetries 
of the singular $V_p$ system. No quantization problem occurs in case 
of the regular spacetime $V_h$. We can see that quantization is very 
sensitive to the choice of spacetime topology.

Generalization of our results to the four dimensional de Sitter spacetimes 
is straightforward. The quantum dynamics of a particle on hyperboloid 
is presented in [8]. The de Sitter spacetime with the topology $R^1\times R^3$, 
the four dimensional analog of $V_p$, is geodesically incomplete and it can 
be embedded isometrically [3] into the four dimensional analog of $V_h$ 
by generalization of the map (3). This is why a direct application of our 
method should lead, after tedious calculations, to the results similar 
in its essence to the results presented here.

We believe that one can generalize our results further to any spacetime with
topology admitting removable type singularities. Quantization of dynamics of 
a test particle in such singular spacetime should be feasible, unless the 
system has no globally well defined observables. 

Further analyses of the quantum dynamics can be carried out at the level 
of the unitary representations of the symmetry groups. The results will be 
published elesewhere [9].

Our paper concerns \emph{removable} type singularities.  It is 
possible that analyses for spacetimes with \emph{essential} 
type singularities [6] may bring  unexpected results.

\vspace{0.5 cm}

{\bf Acknowledgements}

I am very grateful to G. Jorjadze, A. Trautman and S. L. Woronowicz for 
helpful discussions, and to  the anonymous referee for 
the criticism which helped me to clarify the paper.


\end{document}